\documentclass[a4paper,fleqn,usenatbib]{mnras}

\usepackage{newtxtext,newtxmath}
\usepackage[T1]{fontenc}
\usepackage{ae,aecompl}

\usepackage{graphicx}	% Including figure files
\usepackage{amsmath}	% Advanced maths commands

\usepackage{amssymb}	% Extra maths symbols

\newcommand{\msun}{{\rm M}_{\sun}}

\title[RMS pulse fractions for Swift J0243.6+6124]{\textit{Insight-HXMT} observations of  Swift J0243.6+6124: the evolution of RMS pulse fractions at super-Eddington luminosity}
\author[P. J. Wang et al.]{
P. J. Wang$^{1,2}$\thanks{E-mail: wangpj@ihep.ac.cn (PJW),kongld@ihep.ac.cn (LDK), szhang@ihep.ac.cn (SZ)}, L. D. Kong$^{1,2}$\footnotemark[1], S. Zhang$^{1}$\footnotemark[1], Y. P. Chen$^{1}$, S. N. Zhang$^{1,2}$, J. L. Qu$^{1,2}$, 
L. Ji$^{3}$,   
\newauthor
L. Tao$^{1}$, M. Y. Ge$^{1}$, F. J. Lu$^{1}$
% ...
\\
$^{1}$ Key Laboratory of Particle Astrophysics, Institute of High Energy Physics, Chinese Academy of Sciences, Beijing 100049, China\\
$^{2}$ University of Chinese Academy of Sciences, Chinese Academy of Sciences, Beijing 100049, China\\
$^{3}$ Institut f\"ur Astronomie und Astrophysik, Kepler Center for Astro and Particle Physics, Eberhard Karls Universit\"at, 72076 T\"ubingen, Germany\\
$^{4}$ Department of Astronomy, Beijing Normal University, Beijing 100088, China\\
$^{5}$ Department of Astronomy, Tsinghua University, Beijing 100084, China\\
$^{6}$ Department of Physics, Tsinghua University, Beijing 100084, China\\
$^{7}$ Department of Engineering Physics, Tsinghua University, Beijing 100084, China\\
$^{8}$ Key Laboratory of Space Astronomy and Technology, National Astronomical Observatories, Chinese Academy of Sciences, Beijing 100012, China\\
$^{9}$College of physics Sciences \& Technology, Hebei University, Baoding 071002, Hebei Province, China\\
$^{10}$Computing Division, Institute of High Energy Physics, Chinese Academy of Sciences, Beijing 100049, China\\
$^{11}$School of Physics and Optoelectronics, Xiangtan University, Xiangtan 411105, China\\
$^{12}$College of Physics, Jilin University, Changchun 130012, China\\
}
\date{Accepted XXX. Received YYY; in original form ZZZ}
\pubyear{2015}
\begin{document}
\label{firstpage}
\pagerange{\pageref{firstpage}--\pageref{lastpage}}
\maketitle
\begin{abstract}
Based on \textit{Insight-HXMT} data, we report on the pulse fraction evolution  during the 2017-2018 outburst of the newly discovered first Galactic ultraluminous X-ray source (ULX) Swift J0243.6+6124. The pulse fractions of 19 observation pairs selected in the rising and fading phases with similar luminosity are investigated. The results show a general trend of the pulse fraction increasing with luminosity and energy at super-critical luminosity.  However, the relative strength of the pulsation between each pair evolves strongly with luminosity. The pulse fraction  in the  rising phase is larger at luminosity below $7.71\times10^{38}$~erg~s$^{-1}$, but smaller at above. A transition luminosity is found to be energy independent. Such a phenomena is firstly confirmed by \textit{Insight-HXMT} observations and we speculate it may have relation with the radiation pressure dominated accretion disk.\\
\end{abstract}
\begin{keywords}
pulse fraction -- pulse profile -- stars: neutron -- pulsars: individual (Swift J0243.6+6124) -- X-rays: binaries
\end{keywords}
%%%%%%%%%%%%%%%%%%%%%%%%%%%%%%%%%%%%%%%%%%%%%%%%%%
%%%%%%%%%%%%%%%%% BODY OF PAPER %%%%%%%%%%%%%%%%%%
\section{Introduction}
The ultraluminous X-ray sources (ULXs) are pointlike, non-nuclear X-ray sources, whose X-ray luminosities exceed the Eddington limit of Galactic stellar mass black holes (assuming isotropic emission, typically $10^{39}$~erg~s$^{-1}$, BH: ${L}_{\rm Edd}$=$1.3\times10^{38}$$({M}_{\rm BH}/{\msun})$~erg~s$^{-1}$). Some ultraluminous X-ray sources exhibit X-ray pulsations, e.g., M82 X-2 \citep{Bachetti et al.2014}, NGC 7793 P13 \citep{F2016,Israel et al.(2017a)}, NGC 5907 ULX1 \citep{Israel et al.(2017b)}, NGC 300 ULX1 \citep{Carpano et al.(2018)}, NGC 1313 X-2 \citep{Sathyaprakash2019} and M51 ULX-7 \citep{R2019}, which are considered to be ultraluminous X-ray pulsars with luminosity produced by the super-Eddington accretion of the neutron star (NS). 
For the X-ray pulsar, the configuration of emitting region depends on the mass accretion rate. In sub-critical luminosity, the accretion plasma falls directly to the neutron star surface and heats the neutron star atmosphere to generate X-rays. When the source reaches the super-critical luminosity, the accretion material can be decelerated via the radiation dominated shock, and an accretion column will be formed below the shock surface \citep{Basko1976}. The accretion disk may have three distinct zones \citep{Shakura1973}: inner zone (zone A), intermediate zone (zone B), and outer zone (zone C), dominated by radiation pressure, gas pressure and electron scattering, gas pressure and Kramer opacity, respectively. When the luminosity is
high enough, the inner edge of the disk may reach to zone A, i.e., the
disk is dominated by the radiation pressure. The observational signatures of a radiation-pressure dominated accretion disk are reported by \citet{Jilong2019} and \citet{M2019}.
%However, such systems are intrinsic bright but observational faint since they are located in the nearby galaxies.

Swift J0243.6+6124 is the first Galactic ULXs harboring a NS discovered by the Neil Gehrels \textit{Swift} Observatory during the 2017-2018 outburst \citep{Kennea et al.(2017)}. It has a pulse period about 9.86 s and exhibits spin-up during the outburst \citep{Doroshenko et al.(2018), Zhang et al.(2019)}. Its companion star was identified as a Be star \citep{Kouroubatzakis et al.(2017)}, due to the presence of hydrogen and helium emission lines in the optical band. The NS magnetic field strength $B < 10^{13}$ G was estimated by \citet{Tsygankov2018} based on the upper limit on the propeller luminosity ${L}_{\rm prop} < 6.8\times10^{35}$~erg~s$^{-1}$. The timing analysis by \citet{Doroshenko2019} reveals two critical luminosities: a critical luminosity (${L}_{\rm 1}$) of $1.5\times10^{38}$~erg~s$^{-1}$ where the source is expected to be in transition from pencil to fan mode, and a critical luminosity (${L}_{\rm 2}$) of $4.4\times10^{38}$~erg~s$^{-1}$ where the accretion mode of the disk is supposed to change from gas to radiation dominated. \textit{NuSTAR} snapshots show the spectrum can be generally fitted with a cutoff power law plus one blackbody \citep{Jaisawal et al.(2018)} but needs multiple blackbody components at high luminosities \citep{Tao2019}. Part of the latter may be related to the outflow \citep{Eijnden2019}. 

Swift J0243.6+6124 was monitored thoroughly
by \textit{Insight-HXMT} with high cadence and high statistics. In this paper, we report the pulsed fraction evolution of Swift J0243.6+6124 during the 2017-2018 outburst in a broad range of energy obtained by \textit{Insight-HXMT}. The observations and data reduction are described in Section \ref{obser}. In Section \ref{result}, we present the results of pulse fractions and pulse profile analyses. Finally, our arguments to explain the evolution of pulse fractions are discussed in Section \ref{diss}.
\section{Observations and Data analysis}
\label{obser}
\begin{table}
	\begin{center}
		
		\caption{Observational pairs adopted in the pulsed fraction analysis.}
		%\vspace{5pt}
		\begin{tabular}{cccccccccccccccccc}
			\\\hline
			No  & \textit{Insight-HXMT} ObsID	&${L}_{\rm {2-150 keV}}$   & Time \\
			&	&	($10^{38}$~erg~s$^{-1}$) &(MJD)\\
			\hline
			1&P011457700201&3.97&58043.1\\
			&P011457705502&3.92&58100.3\\
			2&P011457700301&4.05&58044.1\\
			&P011457705402&4.10&58099.3\\
			3&P011457700401&4.29&58045.1\\
			&P011457705401&4.29&58099.1\\
			4&P011457700501&4.49&58046.1\\
			&P011457705301&4.47&58098.4\\
			5&P011457700502${\mathrm{*}}$&4.64&58046.3\\
			&P011457704901&4.81&58094.2\\
			6&P011457700502${\mathrm{*}}$&4.64&58046.2\\
			&P011457705101&4.71&58096.6\\
			7&P011457700701&5.59&58049.0\\
			&P011457704501&5.49&58090.0\\
			8&P011457700801${\mathrm{*}}$&5.77&58050.0\\
			&P011457704301&5.89&58088.2\\
			9&P011457700801${\mathrm{*}}$&5.77&58050.0\\
			&P011457704401&5.77&58089.3\\
			10&P011457701105${\mathrm{*}}$&7.56&58052.7\\
			&P011457703501&7.71&58080.0\\
			11&P011457701105${\mathrm{*}}$&7.56&58052.7\\
			&P011457703601&7.40&58081.1\\
			12&P011457701107&8.10&58053.0\\
			&P011457703401&8.30&58079.2\\
			13&P011457701201&9.02&58055.0\\
			&P011457703301&9.08&58078.1\\
			14&P011457701301&9.64&58056.0\\
			&P011457703201&9.45&58077.4\\
			15&P011457701401&10.69&58057.0\\
			&P011457703001&11.01&58075.2\\
			16&P011457701401&10.69&58057.0\\
			&P011457703101&10.42&58076.2\\
			17&P011457701501&11.77&58058.0\\
			&P011457702901&11.97&58074.2\\
			18&P011457701801${\mathrm{*}}$&19.51&58061.3\\
			&P011457702101&20.65&58065.6\\
			19&P011457701801${\mathrm{*}}$&19.51&58061.3\\
			&P011457702102&20.52&58065.8\\
			\hline
			\label{obs-lum}
		\end{tabular}
	\begin{list}{}{}
		\item[${\mathrm{*}}$]{The duplicate observations in different pairs.}
		%\item[Note:]{The OBSID, time,  peak flux  in 2-10 keV (in units of cts/s), PRE or non-PRE bursts  are given.}
	\end{list}
	\end{center}
	
\end{table}

\begin{figure}
	\centering
	\includegraphics[angle=0, scale=0.35]{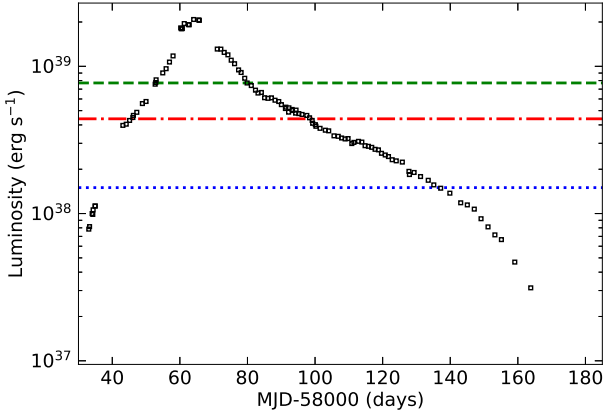}
	\caption{ The luminosity evolution of Swift J0243.6+6124 as observed by \textit{Insight-HXMT} during outburst. The green dashed stands for the luminosity discovered with ${f}_{\rm rms}$ transition. The blue dotted line and red dash-dot line represent ${L}_{\rm 1}$ and ${L}_{\rm 2}$ as reported by \citet{Doroshenko2019}, respectively. 
	}
	\label{lum}
\end{figure}

\textit{Insight-HXMT} is the first Chinese X-ray satellite and it was launched on June 15th 2017. \textit{Insight-HXMT} was designed in a collimated mode and composed of three telescopes: the High Energy X-ray telescope (HE, poshwich NaI/CsI, 20--250 keV),  the Medium Energy X-ray telescope (ME, Si pin detector, 5--30 keV) and the Low Energy X-ray telescope (LE, SCD detector, 0.7--13 keV), working in scanning and pointing observational modes and GRB mode. See details about \textit{Insight-HXMT} in \citet{Zhang2019, Cao2019,Liu2019,Chen2019}. \textit{Insight-HXMT} observations of Swift J0243.6+6124 cover a period from MJD 58030 to MJD 58180 (Figure \ref{lum}). To investigate the pulse properties in the rising and fading phases with a similar luminosity, as shown in Figure \ref{lum}, we select the \textit{Insight-HXMT} observational pairs available within MJD 58043--58100, when the source stayed at super-Eddington luminosities. This paper focuses on the timing analysis of the pulse fraction, the luminosity estimation follows the procedure in \citet{Zhang et al.(2019)}, where the spectral model is 
{\tt cons$*$TBabs$*$(cutoffpl+bbody+gaussian)}. The luminosity is obtained in 2--150 keV for Swift J0243.6+6124 by assuming a distance of 6.8 kpc \citep{Bailer-Jones2018}.  A systematic error of 1\% is added in the spectral fitting with the XSPEC software package version 12.10.0c.
The resulted luminosities are compared and those pairs  with luminosity consistency within 97--103\% are selected. Finally, a sample of 19 pairs is obtained which consists of 14 observations in the rising part and 19 observations in the fading part, as is shown in Table \ref{obs-lum}. For the timing analysis, \textit{Insight-HXMT} data analysis software package {\tt HXMTDAS V2.01} is used. \textit{Insight-HXMT} data are processed according to the standard processing procedures described in the \textit{Insight-HXMT} Data Reduction Guide v2.01\footnote[1]{{http://www.hxmt.org/index.php/usersp/dataan/fxwd}}. 
The arrival time of each photon is corrected to the solar system barycenter with the {\tt HXMTDAS} tool {\tt hxbary} and corrected for the orbital modulation by taking the ephemeris coming from GBM Pulsar Spin Histories\footnote[2]{{https://gammaray.msfc.nasa.gov/gbm/science/pulsars.html}}.
By using the \textit{Stingray}\footnote[3]{{https://stingray.readthedocs.io/en/latest/}} model in \textit{python}, the spin period of Swift J0243.6+6124 is measured in each observation. In the timing analysis the entire \textit{Insight-HXMT} band is subdivided into 12 energy bins, and the details  are shown in Table \ref{hxmt-en}. Finally, pulse fractions and profiles are obtained in each energy bin of the observational pairs.

\begin{table*}
	\begin{center}
		
		\caption{The energy bins of each Insight-HXMT telescopes used for pulse fraction analysis. }
		%\vspace{5pt}
		\begin{tabular}{cccccccccccccccccc}
			\\\hline
			Telescopes of \textit{Insight-HXMT}  & ${E}_{\rm 1}$ &${E}_{\rm 2}$  & ${E}_{\rm 3}$ &${E}_{\rm 4}$\\\hline
			LE&0.8-2.4 keV&2.4-3.4 keV&3.4-4.4 keV&4.4-5.6 keV\\
			ME&6.5-9.8 keV&9.8-12.9 keV&12.9-17.7 keV&17.7-28.2 keV\\
			HE&27.4-32.4 keV&32.4-38.6 keV&38.6-51.6 keV&51.6-107.9 keV\\
			
			\hline
			\label{hxmt-en}
		\end{tabular}
	\end{center}
	
\end{table*}

\section{Results}
\label{result}

\subsection{Energy dependence of the pulse fraction}
\label{section-frms-en}

%==============================frms-en=================================
\begin{figure}
	\centering
	\includegraphics[angle=0, scale=0.35] {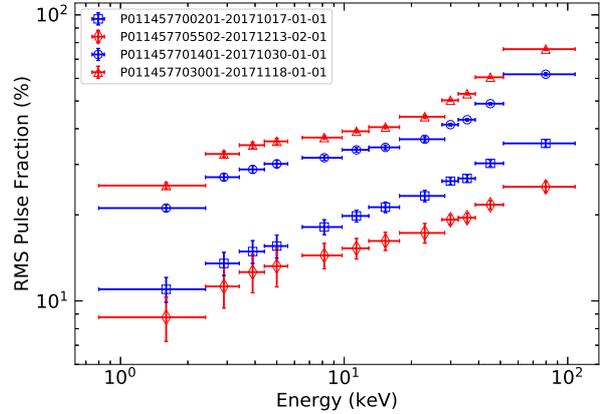}
	\caption{Energy spectra of the RMS pulse fraction derived for two pair observations. The data with circle and triangle (No. 1) have larger luminosity than those with square and diamond (No. 15). The rising phases are denoted by square and circle, and the fading phases by diamond and triangle.  
	}
	\label{frms-en}	
\end{figure}
%The diamond present the luminosity of the pair of observations below the characteristic luminosity.

\begin{figure}
	\centering
	\includegraphics[angle=0, scale=0.35] {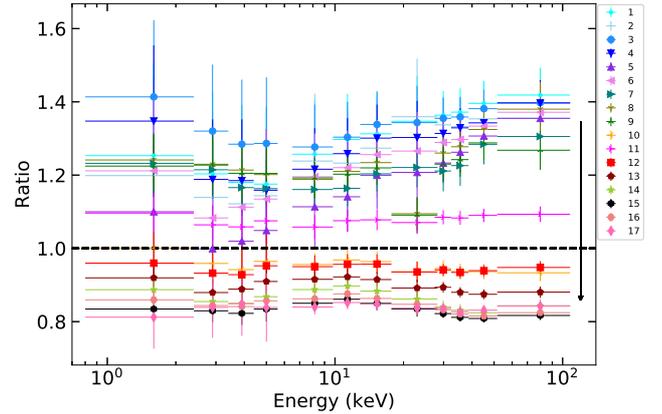}
	\caption{Energy spectra of the pulse fraction ratio between the rising and fading phases for 17 observational pairs. The vertical line with arrow shows the direction with the increasing luminosity. Two observational pairs with the highest luminosity are not included because they were carried out closely.
	}
	\label{ratio-en}
\end{figure}

\begin{figure}
	\centering
	\includegraphics[angle=0, scale=0.35] {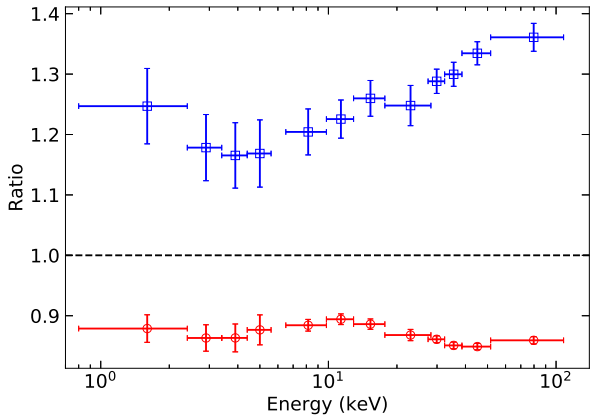}
	\caption{Same as in Figure \ref{ratio-en}, but average over observations with ratio above 1 (blue square) and below 1 (red circle).
	}
	\label{two-distribution}
\end{figure}

%===============================lum-frms=============================

The root mean squared (RMS) for the pulsation, denoted as ${f}_{\rm rms}$, are computed for each energy bin of the entire 33 observations. Here the ${f}_{\rm rms}$ is defined as:

\begin{equation}
%\mathbf{F}_{\rm rms} = \frac{(\sum\nolimits_{i=1}^N(r_i-\bar{r})^2 /N)^{1/2}}{\bar{r}}
{f}_{\rm rms} = \frac{(\sum\nolimits_{i=1}^N(r_i-\bar{r})^2 /N)^{1/2}}{\bar{r}}
\end{equation}
where $\bar{r}$ is the phase average count rate, $r_i$ the phase count rate and $N = 32$ the total phase bin number. The error of ${f}_{\rm rms}$ is estimated by error transfer and quoted at the 90\% confidence level.
We find that in all the observations ${f}_{\rm rms}$ increases with energy, which is consistent with that obtained with a simple definition of pulse fraction ($(F_{\rm max}-F_{\rm min})/(F_{\rm max}+F_{\rm min})$) reported by \citet{Tao2019} with \textit{NuSTAR} data.
An example of such an energy dependence on ${f}_{\rm rms}$ for two pairs with a relatively lower luminosity ($\sim 3.95\times10^{38}$~erg~s$^{-1}$) and a higher luminosity ($\sim 1.09\times10^{39}$~erg~s$^{-1}$) is shown in Figure \ref{frms-en}, where the energy dependence of the pulse fraction is obvious. The pulse fraction of the rising phase is larger at lower luminosity but smaller at higher luminosity than that of the fading phase. Such a trend is seen more clearly in Figure \ref{ratio-en} where the ${f}_{\rm rms}$ ratio of  the rising to the fading  observations of each pair is plotted against energy. Along with the increasing luminosity, such a ratio moves from > 1 toward <1.  The ratios averaged over each energy bin for data with ratio above and below a unit, as shown in Figure \ref{two-distribution}, clearly show the distinct evolution behavior. It seems  that, ${f}_{\rm rms}$ of the rising and fading phases tends to be comparable at around a specific luminosity. Such a trend is indicated in \textit{NICER} observations at energies below 12 keV \citep{Wilson-Hodge et al.(2018)}. We hence investigate this in what follows via looking into the luminosity dependence in each energy bin.

\subsection{Luminosity dependence of the pulse fraction }
\label{evolution-frms-lum}
%====================LE,ME,HE=======================================

\begin{figure}
	\centering
	\includegraphics[angle=0, scale=0.32] {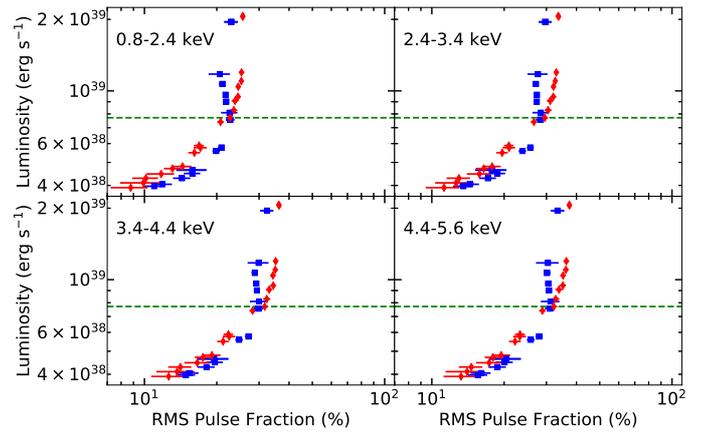}
	\caption{${f}_{\rm rms}$ evolution against luminosity for each energy bin. The green dashed represents the cross luminosity when ${f}_{\rm rms}$ of the rising and fading phases are comparable. The rising and fading phases are denoted by
		blue square and red diamond, respectively.
	}
	\label{lum-frms_LE}
\end{figure}

\begin{figure}
	\centering
	\includegraphics[angle=0, scale=0.32] {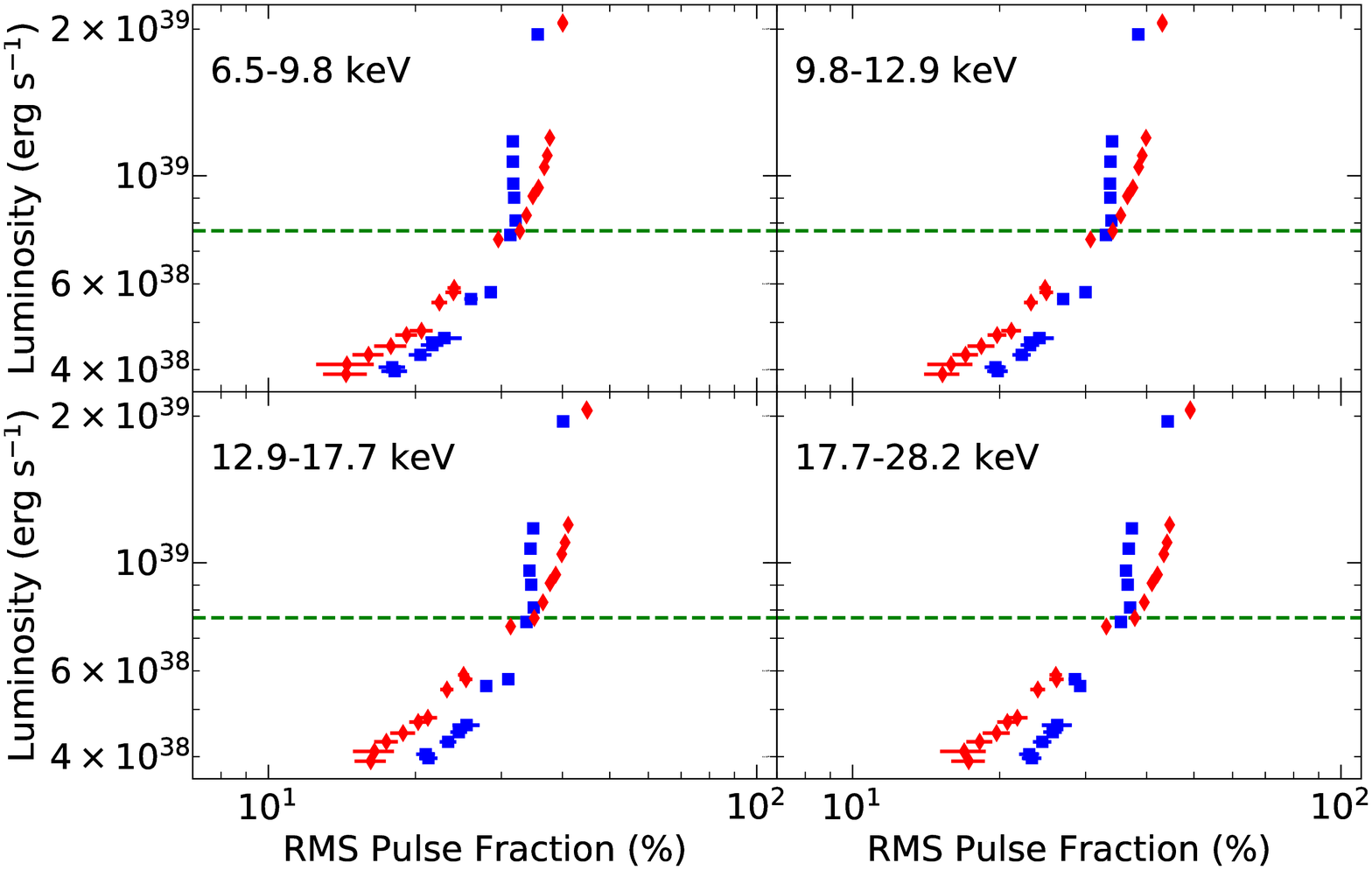}
	\caption{Same as Figure \ref{lum-frms_LE} but with different energy bins.
	}
	\label{lum-frms_ME}
\end{figure}

\begin{figure}
	\centering
	\includegraphics[angle=0, scale=0.32] {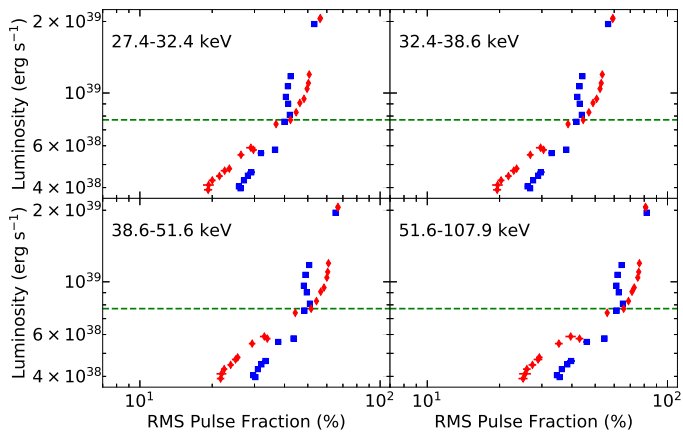}
	\caption{Same as Figure \ref{lum-frms_LE} but with different energy bins.
	}
	\label{lum-frms_HE}
\end{figure}
%===========================================================================

\begin{figure}
	\centering
	\includegraphics[angle=0, scale=0.35] {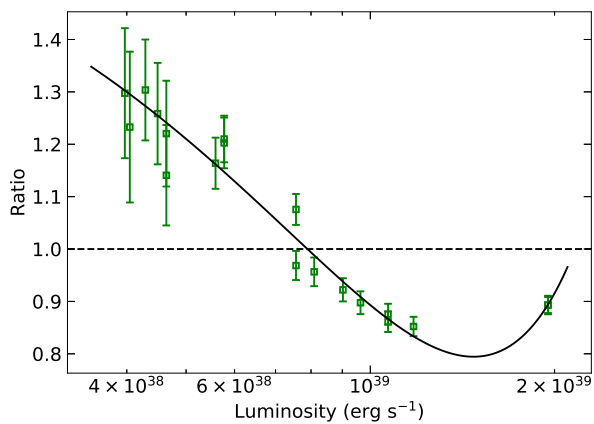}
	\caption{The ${f}_{\rm rms}$ ratio evolution against luminosity in 9.8--12.9 keV. The green square and black line represent the data and the quadratic polynomial fitting, respectively. The black dash represents where the ratio is equal to unit.
	}
	\label{ratio-fit}
\end{figure}

\begin{figure}
	\centering
	\includegraphics[angle=0, scale=0.28] {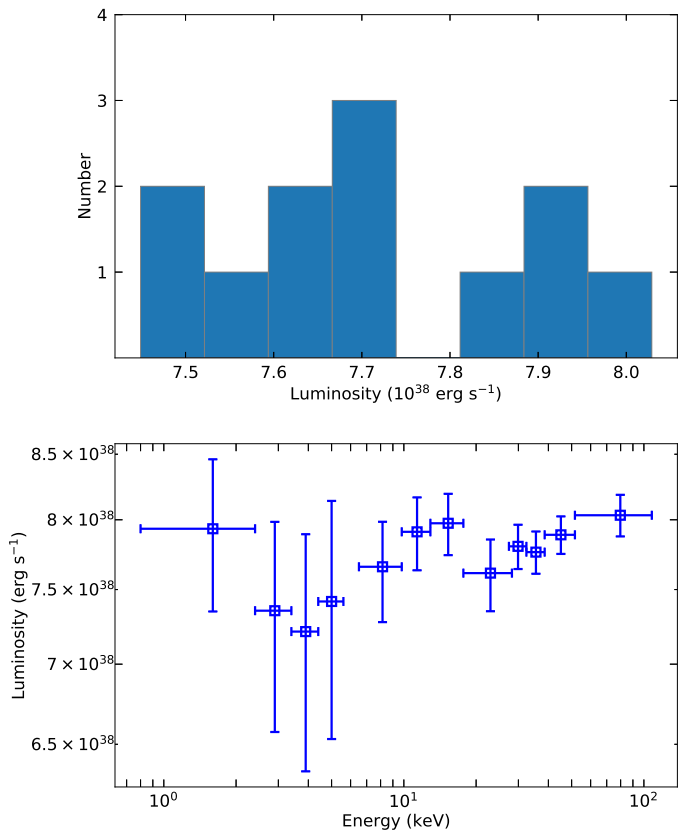}
	\caption{Upper panel: the distribution of the cross luminosity obtained by quadratic polynomial fitting in 12 energy bins. Lower panel: the cross luminosity versus energy. The mean cross luminosity is $7.71\times10^{38}$~erg~s$^{-1}$.
	}
	\label{find_value}
\end{figure}

In Figures \ref{lum-frms_LE}, \ref{lum-frms_ME}, \ref{lum-frms_HE}, ${f}_{\rm rms}$ for each pair are plotted in each of the 12 energy bins.  One sees that, along with the increasing luminosity, for all the energy bins, ${f}_{\rm rms}$ evolves in a similar way in each pair: ${f}_{\rm rms}$ is larger at lower luminosity and then gradually approaches to that in the fading phase, and finally the trend reverses once the luminosity passes through a certain value. To find out the energy dependence of the cross luminosity, we plot in Figure \ref{ratio-fit} the ${f}_{\rm rms}$ ratios of each pair in an energy bin of 9.8--12.9 keV. In order to obtain
the range of the cross luminosity, we use the quadratic polynomial to fit the ${f}_{\rm rms}$ ratio, as shown in Figure \ref{ratio-fit}. To investigate the energy dependence of the cross luminosity, the data in Figure \ref{ratio-fit} are re-sampled via bootstrapping for error estimations. As shown in Figure \ref{find_value} are the luminosity distribution and the energy evolution of the luminosity obtained via quadratic polynomial fittings. The histogram distribution gives a mean cross luminosity of $7.72\times10^{38}$~erg~s$^{-1}$, which is well consistent with that of $7.71_{-0.14}^{+0.12}\times10^{38}$~erg~s$^{-1}$, obtained by averaging over the adopted energies. One sees from Figure 9 that the cross luminosity is most likely energy independent.

%=================================find-lum============================

%========================================================================== 

\subsection{Evolution of the pulse profile}

Nineteen pairs of pulse profiles are presented in Figures \ref{pro_0.8}--\ref{pro_38.6}, with adjacent energies combined due to their similarity in pulse profile. Each pulse profile is normalized to its mean count rate. Pulse profiles for each pair are co-aligned with respect to the phase with the minimum count rate. From these plots, we see that the pulse profiles are in general similar for each pair between the rising and the fading phases. Along with the increasing luminosity, the peak flux of the pulses in  the rising phase decreases and finally becomes smaller than those in the fading phase once the source is brighter than the cross luminosity $7.71\times10^{38}$~erg~s$^{-1}$.

\section{Discussion and Summary}
\label{diss}

\begin{figure*}
	\centering
	\includegraphics[angle=0, scale=0.2] {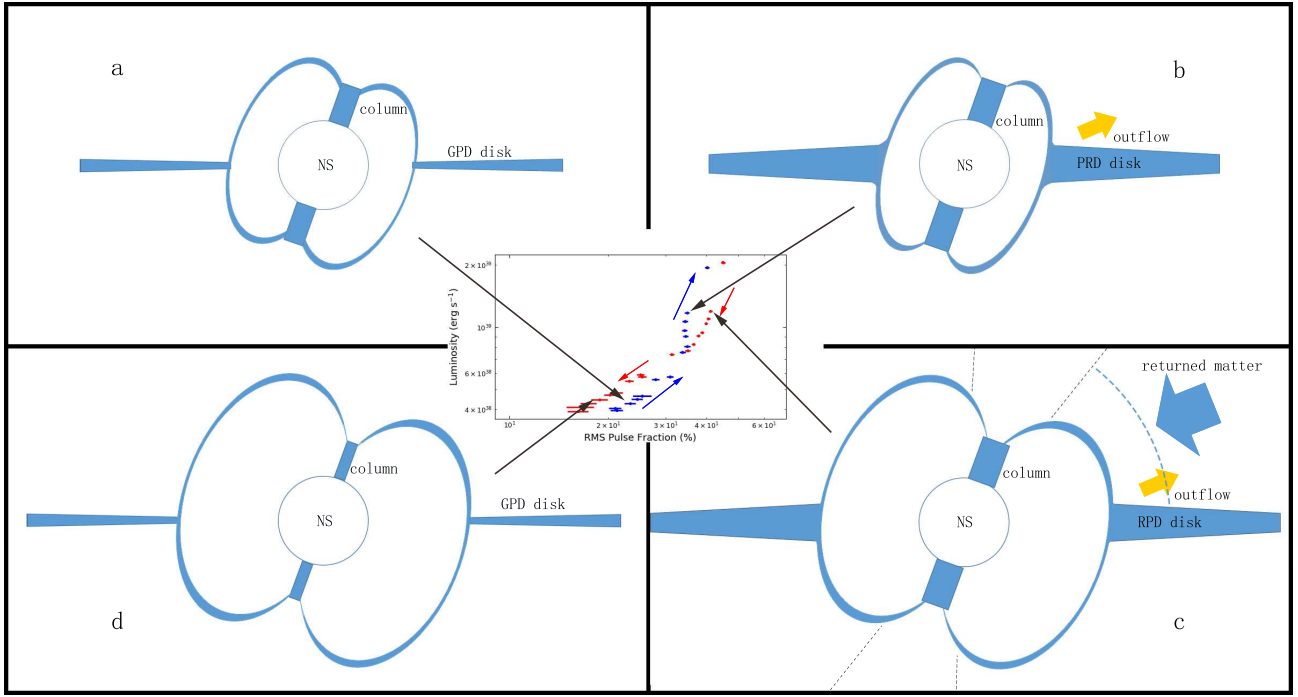}
	%\caption{0.8-2.4kev}
	\caption{Schematic drawing of the evolution of the accretion disk and accretion column, the latter produces the observed pulsed-emission. The inset in the middle shows the observed evolution of the RMS pulse fraction (${f}_{\rm rms}$) during the rising phase (blue crosses and arrows) and fading phase (red crosses and arrow).  Panels a, b, c and d show the proposed states of the accretion disk and accretion column during the rising phase (a and b) and fading phase (c and d), respectively. In panel a, ${L}$<${L}_{\rm 2}$, the disk is gas pressure dominated. In panel b, ${L}$>${L}_{\rm 2}$, the disk is radiation pressure dominated and outflow from the disk is produced; ${f}_{\rm rms}$ increases slightly due to the slightly broader accretion column caused by the smaller magnetospheric radius at higher accretion rate. In panel c, some of the outflowed matter is returning back to the system, following a broader range of magnetic field lines to form a broadened accretion column, thus producing higher ${f}_{\rm rms}$ for the same luminosity compared to that in panel b. In panel d, the disk is back to gas pressure dominated state and ${f}_{\rm rms}$ is reduced due to the larger magnetospheric radius than that in panel a for the same luminosity, if the same physical process happened as that observed in the HMXB V 0332+53 \citep{Doroshenko2017}. }
	\label{ps}
\end{figure*}

The high-cadence \textit{Insight-HXMT} observations of the first Galactic ULX Swift J0243.6+6124 allow us for the first time to compare the pulse properties of the rising and fading phases in detail at similar luminosities. We select 19 pairs of observations covering the outburst evolution with luminosities above ${L}_{\rm 1}$. Previously such researches were relatively rare due to sporadic coverage of HMXB outbursts and faintness of the ULXs located in the neighboring galaxies. For Swift J0243.6+6124, although the snapshots of \textit{NuSTAR} show an evolving RMS pulse fraction spectrum, and also the different RMS pulse fraction between rising and fading phase is hinted in \textit{NICER} observations at soft X-rays, details were still lacking, especially in a broad energy band for a well-sampled outburst \citep{Tao2019, Wilson-Hodge et al.(2018)}. Our results show that the RMS pulse fraction spectrum is consistent in general with that reported by \textit{NuSTAR} but ends up with the discovery that the relative strength between the rising and fading phases evolves with luminosity: above ${L}_{\rm 2}$, the RMS pulse fraction of the rising phase is in general larger than that in fading phase and, along with the increasing luminosity, such a trend inverses. Further investigations of such a behavior on its energy and luminosity dependences show that the turnaround of RMS pulse fraction occurs at a luminosity around $7.71\times10^{38}$~erg~s$^{-1}$ which has little dependence of energy. The pulse profiles of the rising/fading phase are similar in each pair but the evolving pulse profile peak may provide hint to possible understanding of such the peculiar RMS pulse fraction behavior. 

It is generally believed that for a HMXB the matter from the companion can be accreted towards the compact star via stellar wind. In case of harboring a magnetized neutron star, the accretion rate must be high enough to overcome the magnetic barrier of so-called propeller effect so that the accreting matter can be channeled via magnetic field line onto the magnetic poles of neutron star, where part of the gravitational energy is released in form of X-ray emissions. Depending on the accretion rate, the accreted matter will form a mound or a column at the magnetic poles, and pencil or fan beam modes are expected at work respectively. The critical luminosity for transition between these two modes is estimated as
%\begin{equation}
$L^* \approx \frac{c}{\kappa_{\rm eff}}{l}_{\rm 0} \frac{GM}{R}$ \citep{Mushtukov2015a},
%\end{equation}
here ${\kappa_{\rm eff}}$, ${l}_{\rm 0}$, ${M}$ and ${R}$ are the effective opacity, the annular arc thickness, the NS mass and radius.
The X-rays are most likely to be emitted radially along the field line in pencil mode and perpendicular to the field line in the fan mode. For Swift J0243.6+6124, this luminosity appears as a critical luminosity of $1.5\times10^{38}$~erg~s$^{-1}$. Therefore the emission pattern for the selected 19 pairs is most likely in the fan mode. In the fan mode, the accretion matter will be shocked via radiation pressure from the hot spot on the surface of the neutron star and form a column structure with bulk velocity larger at the top of column \citep{Wang1981, Basko1976, Becker2012}. The seed photons from either the surface of neutron star or the accretion disk will be up-scattered via inverse Compton process in this column region and escape in direction perpendicular to the falling flow. Above ${L}_{\rm 2}$, the accretion mode will switch from a gas dominated into a radiation dominated disk, and hence the inner accretion disk inflates in direction vertical to the disk plane \citep{Doroshenko2019}. 

As shown in a simple definition of pulse fraction ($(F_{\rm max}-F_{\rm min})/(F_{\rm max}+F_{\rm min})$), the pulse fraction is mostly determined by the maximum and minimum fluxes (${F}_{\rm max}$ and ${F}_{\rm min}$) showing up in the pulse profile. The pulse fraction increases with larger ${F}_{\rm max}$ or smaller ${F}_{\rm min}$. In fan mode the pulse fraction of Swift J0243.6+6124 is observed to be proportional to energy and luminosity. The maximum flux in fan mode can be estimated as:
\begin{equation}
L^{**} \!\! = L (H=R) \approx 1.8\times 10^{39}  \!\!  \left( \!\frac{l_0/d_0}{50} \!\! \right)\left(\frac{\kappa_{\rm T}}{\kappa_{\perp}}\right) \frac{M}{\msun}{\rm erg\,s^{-1}} 
\end{equation}
where ${l}_{\rm 0}$ and ${d}_{\rm 0}$ is the annular arc length and thickness of the accretion channel \citep{Mushtukov2015b}.
One sees that the flux increases with a broader column (e.g. with larger ${l}_{\rm 0}$). Along with the increasing luminosity, the accretion disk moves inward and can cover more magnetic field lines which can result in a broadened accretion column on the magnetic pole of the neutron star. The energy dependence of the pulse fraction may be related to the beaming effect. In the accretion column, the accretion material passing through shock will still have relativistic speed at the top of the accretion column where the seed photons are up-scatted to  higher energies, in their way of sinking toward the neutron star surface. Accordingly, the emitted harder X-rays will undergo a larger beaming effect. One has a minimum flux when the orientation of the accretion column aligns with the line of the sight. Such a beaming effect may result in smaller minimum flux for harder X-rays.

During the outburst of the HMXB V 0332+53, it was observed that the magnetospheric radius of the neutron star tends to be smaller (for the same field strength of
the NS) in the rising phase than in the fading phase \citep{Doroshenko2017}. If a similar case holds as well for Swift J0243.6+6124, we speculate that the following scenario may be relevant to the observed transition of the pulse fraction between the rising and fading phases. For the pulse fraction evolution at the luminosity below ${L}_{\rm 2}$, with a smaller magnetospheric radius at the rising phase the accretion disk can move to an inner region, and hence form a broader accretion column at the magnetic pole. Above ${L}_{\rm 2}$, the accretion disk is radiation pressure dominated. 
Part of the accreting material will be blown out but some of them may be bound by gravity and return along a wide range of magnetic field lines to form a broader accretion column during the fading phase.
So the observed transition luminosity of $\sim 7.71\times10^{38}$~erg~s$^{-1}$ may be the result of the balance between these two effects, which is evolving as the outburst luminosity goes beyond ${L}_{\rm 1}$ (Figure \ref{ps}).  

In summary, \textit{Insight-HXMT} observed a peculiar evolution of the pulse fractions from the first Galactic ULX Swift J0243.6+6124. The balance of the pulse fraction between the rising and fading phases evolves with luminosity: the pulse fraction in the rising phase is larger below a critical luminosity of $7.71\times10^{38}$~erg~s$^{-1}$ but smaller at above. Such a phenomena is firstly confirmed by \textit{Insight-HXMT} observations but a thorough scenario is still missing, although we speculate it may have relation with the transition of the accretion modes between a gas and a radiation pressure dominated disk.

\section*{Acknowledgements}

This work is supported by the National Key R\&D Program of China (2016YFA0400800) and the National Natural Science Foundation of China under grants U1838201, U1838202, U1938101 and 11733009. This work made use of data from the \textit{Insight-HXMT} mission, a project funded by China National Space Administration (CNSA) and the Chinese Academy of Sciences (CAS).

%%%%%%%%%%%%%%%%%%%%%%%%%%%%%%%%%%%%%%%%%%%%%%%%%%

\section*{Data Availability}

The data underlying this article will be shared on reasonable request to the corresponding author.



%\bibliographystyle{mnras}


%%%%%%%%%%%%%%%%%%%%%%%%%%%%%%%%%%%%%%%%%%%%%%%%%%

%%%%%%%%%%%%%%%%% APPENDICES %%%%%%%%%%%%%%%%%%%%%

\appendix
\section{Pulse profiles in six energy bands}
%=================================pulse profile================================
\begin{figure*}
	\centering
	\includegraphics[angle=0, scale=0.27] {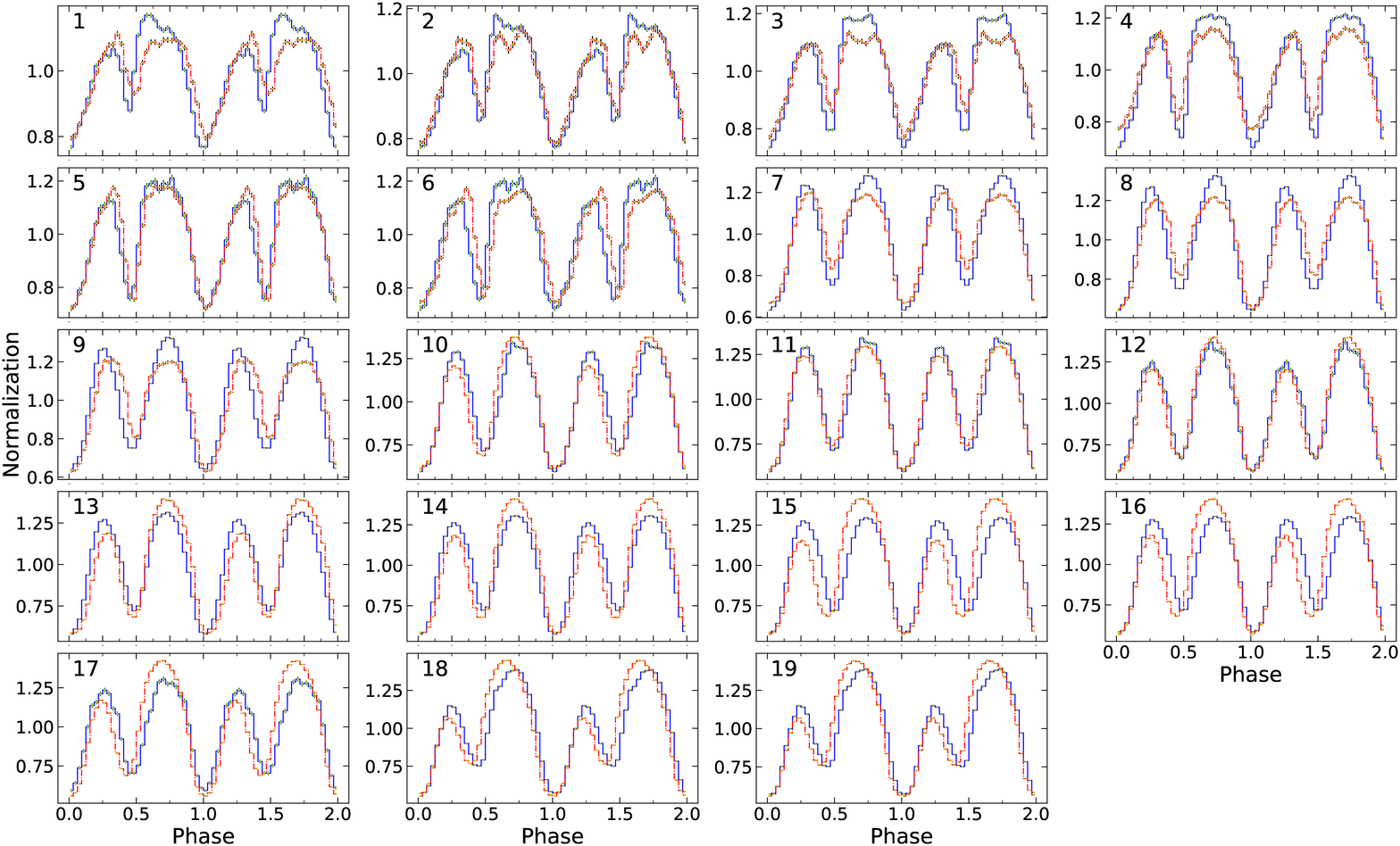}
	%\caption{0.8-2.4kev}
	\caption{Pulse profiles for 19 pairs in 0.8--3.4 keV. The blue line and red dash-dot line represent the rising and fading phases, respectively. For each plot the pulse profile is normalized to their mean value and co-aligned  with respect to the minimum of the pulse profile.}
	\label{pro_0.8}
\end{figure*}

\begin{figure*}
	\centering
	\includegraphics[angle=0, scale=0.27]{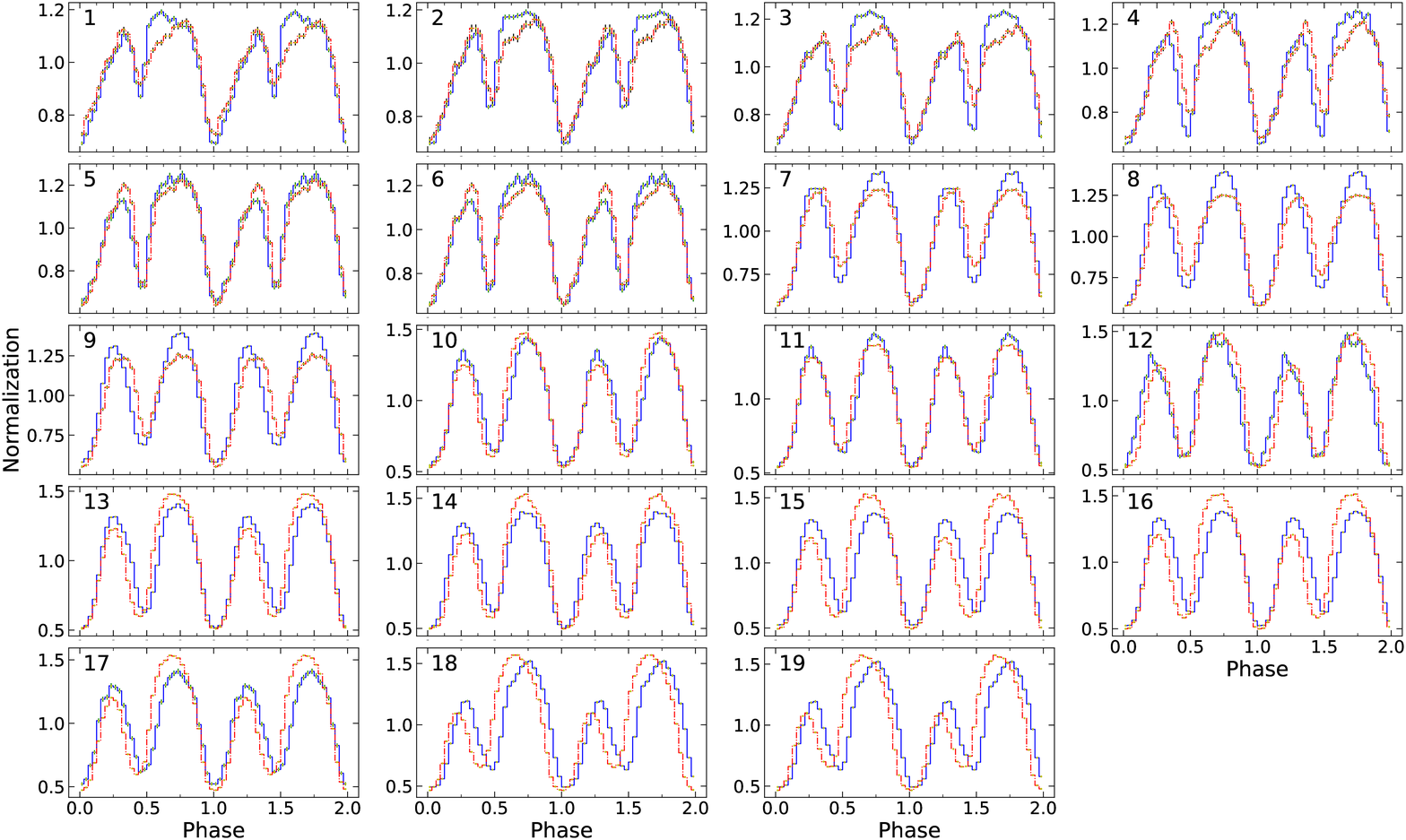}
	\caption{Same as Figure \ref{pro_0.8} but in 3.4--5.6 keV.
	}
	\label{pro_3.4}
\end{figure*}

\begin{figure*}
	\centering
	\includegraphics[angle=0, scale=0.27] {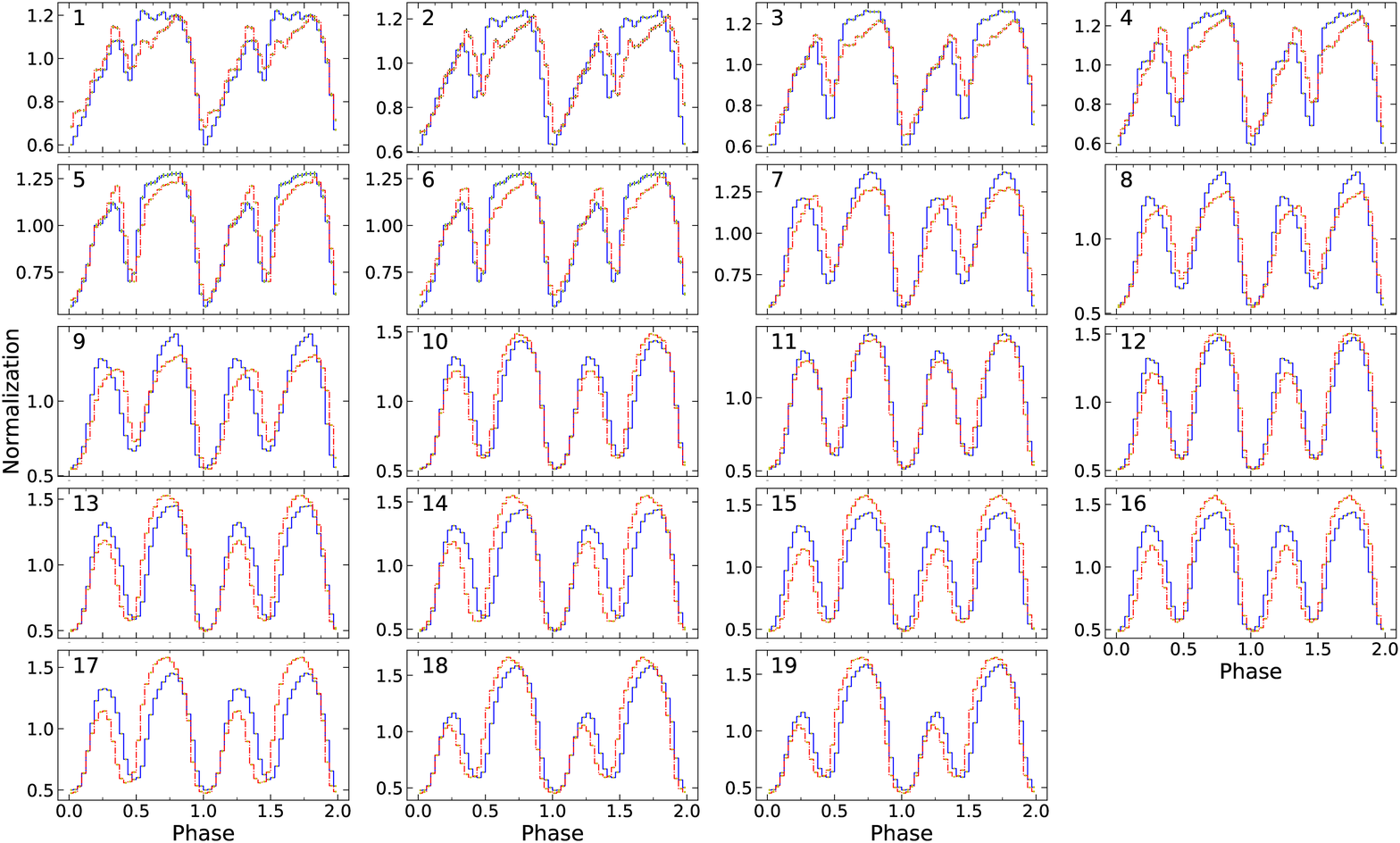}
	\caption{Same as Figure \ref{pro_0.8} but in 6.5--12.9 keV.
	}
	\label{pro_6.5}
\end{figure*}

\begin{figure*}
	\centering
	\includegraphics[angle=0, scale=0.27] {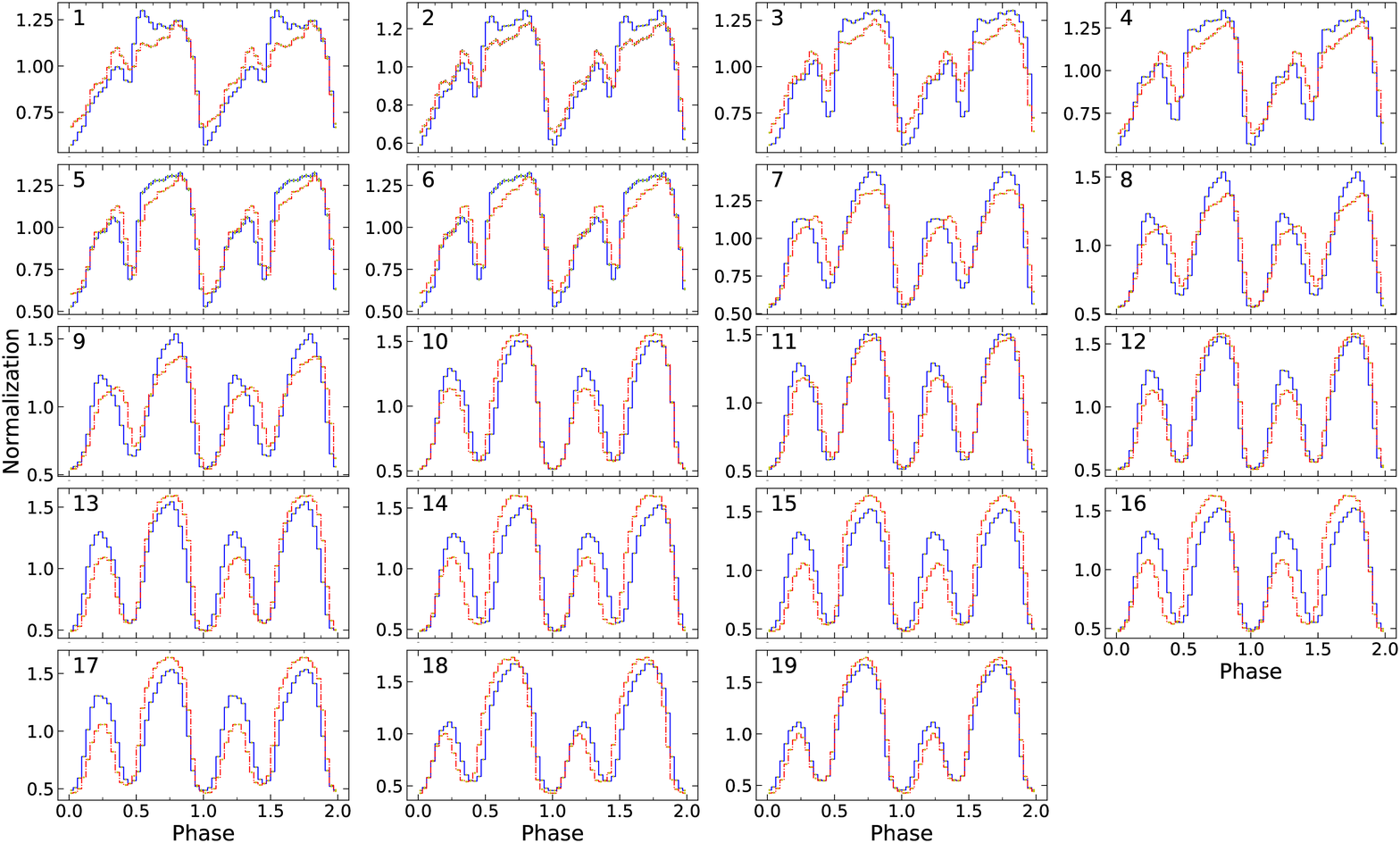}
	\caption{Same as Figure \ref{pro_0.8} but in 12.9--28.2 keV.
	}
	\label{pro_12.9}
\end{figure*}

\begin{figure*}
	\centering
	\includegraphics[angle=0, scale=0.27] {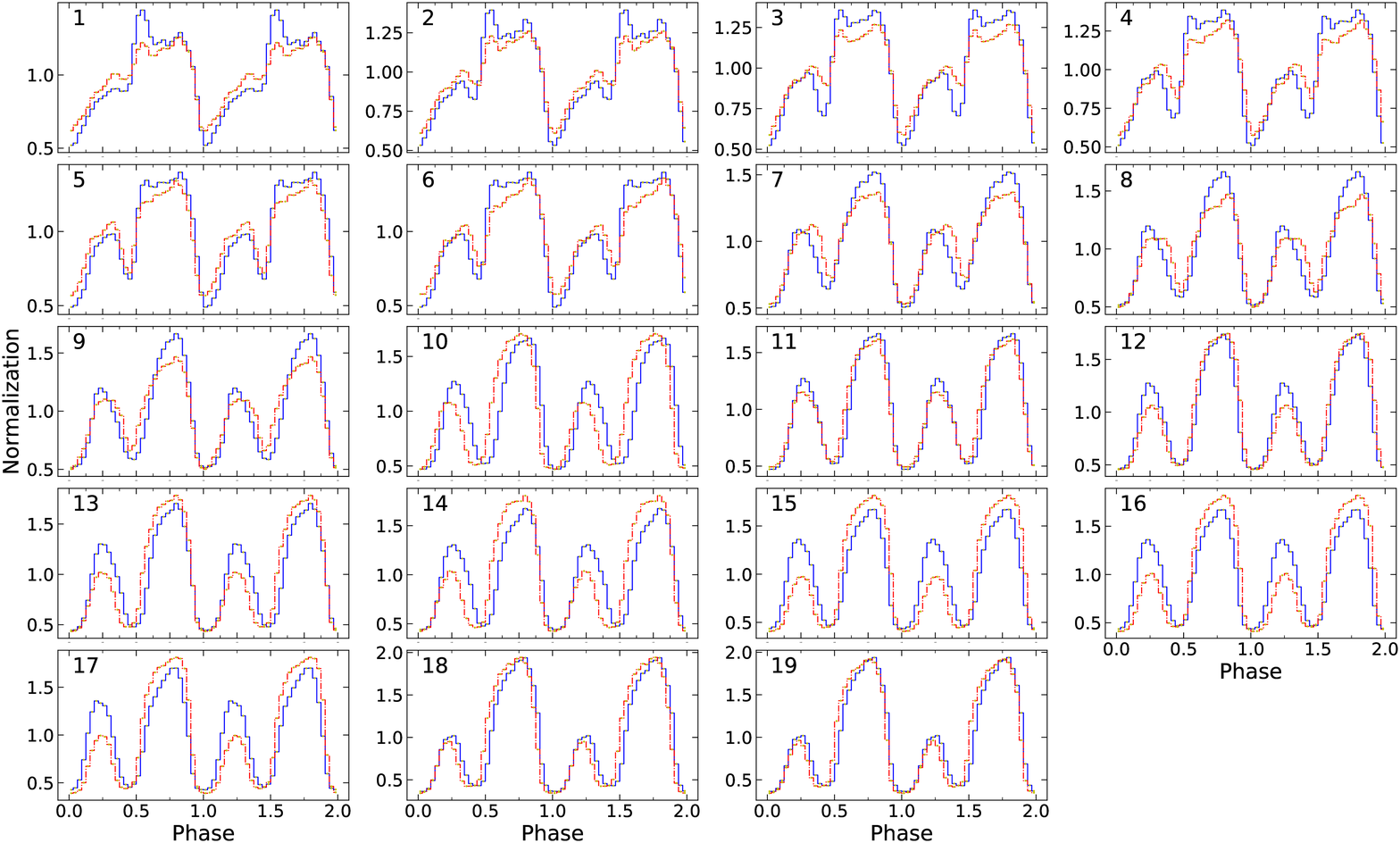}
	\caption{Same as Figure \ref{pro_0.8} but in 27.4--38.6 keV.
	}
	\label{pro_27.4}
\end{figure*}

\begin{figure*}
	\centering
	\includegraphics[angle=0, scale=0.27] {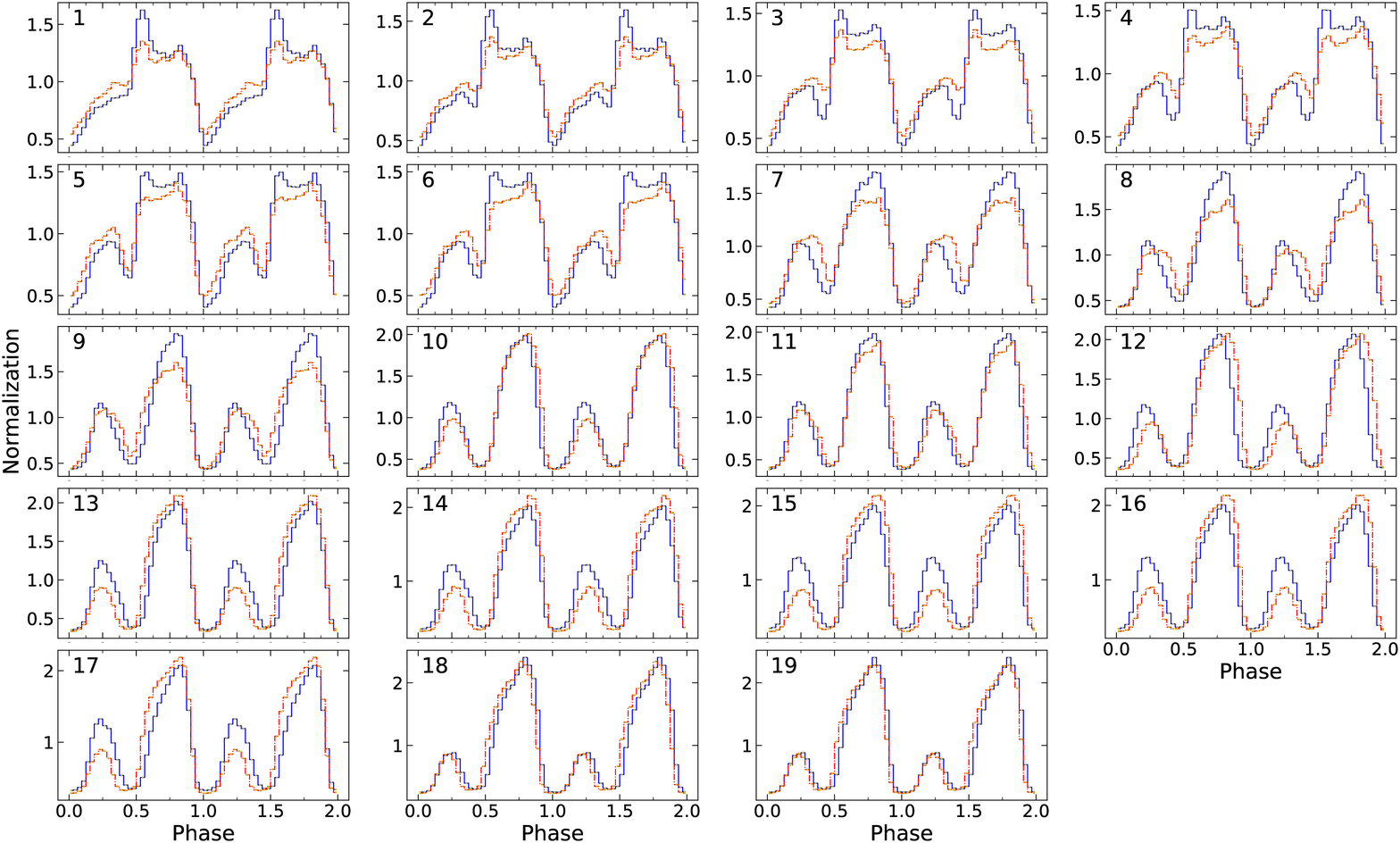}
	\caption{Same as Figure \ref{pro_0.8} but in 38.6--107.9 keV.
	}
	\label{pro_38.6}
\end{figure*}

%%%%%%%%%%%%%%%%%%%%%%%%%%%%%%%%%%%%%%%%%%%%%%%%%%

% Don't change these lines
\bsp	% typesetting comment
\label{lastpage}
\end{document}